\documentclass[twocolumn,aps,prd,amsmath,amssymb]{revtex4}
\bibpunct{}{}{,}{s}{}{,}

\usepackage{bm}
\usepackage{amsmath}
\usepackage{amssymb}
\usepackage{latexsym}
\usepackage{amsfonts}
\usepackage{epsfig}
\usepackage{psfrag}
\usepackage{graphicx}

\newcommand{\ket}[1]{\left|#1\right>}
\newcommand{\bra}[1]{\left<#1\right|}

\newcommand{\f}[1]{\mbox{\boldmath$#1$}}
\newcommand{\fk}[1]{\mbox{\boldmath$\scriptstyle#1$}}
\newcommand{\vau}{\mbox{\boldmath$v$}}
\newcommand{\na}{\mbox{\boldmath$\nabla$}}
\newcommand{\bea}{\begin{eqnarray}}
\newcommand{\ea}{\end{eqnarray}}
\newcommand{\eea}{\end{eqnarray}}
\newcommand{\ord}{{\cal O}}
\newcommand{\sumint}[1]
{\begin{array}{c}
\\
{{\textstyle\sum}\hspace{-1.1em}{\displaystyle\int}}\\
{\scriptstyle{#1}}
\end{array}}

\begin{document}

\title{Recreating Fundamental Effects in the Laboratory?}

\author{Ralf Sch\"utzhold}

\email{ralf.schuetzhold@uni-due.de}

\affiliation{Fachbereich Physik, 
Universit\"at Duisburg-Essen, 
D-47048 Duisburg, Germany}

\begin{abstract}
This article provides a brief (non-exhaustive) overview of some possibilities 
for recreating fundamental effects which are relevant for black holes 
(and other gravitational scenarios) in the laboratory.
Via suitable condensed matter analogues and other laboratory systems, 
it might be possible to model the Penrose process (superradiant scattering),
the Unruh effect, Hawking radiation, the Eardley instability, 
black-hole lasers, cosmological particle creation, the Gibbons-Hawking effect,
and the Schwinger mechanism.
Apart from an experimental  verification of these yet unobserved phenomena,
the study of these laboratory systems might shed light onto the underlying
ideas and problems and should therefore be interesting from a (quantum)
gravity point of view as well. 

\bigskip

\end{abstract}

\maketitle


\section{Introduction}

Under the influence of extreme conditions, such as the strong gravitational 
field around black holes, matter -- and even the vacuum -- behaves in 
unexpected ways and shows many fascinating phenomena \cite{Birrell}
like Hawking radiation \cite{Hawking}.
Although extremely hard to observe, these striking effects are very 
interesting form a fundamental point of view for our understanding 
of black holes and (quantum) gravity.
For example, Hawkings discovery of black hole evaporation \cite{Hawking} 
surprisingly confirmed Bekensteins thermodynamic interpretation 
\cite{thermo} of black holes.
Understanding this link to thermodynamics including the origin of the 
black hole entropy and the related black hole information ``paradox'' 
is one of the key questions of quantum gravity.

Since the options for an experimental or observational approach to these 
problems are rather limited, an alternative possibility would be to 
recreate the aforementioned phenomena in suitable laboratory settings
\cite{Living,Droplet,Novello,LNP,Memorial}.
This idea will be pursued in the following.
After a short review of the fundamental effects under consideration in 
Sec.~\ref{Fundamental}, the underlying idea of analogue gravity is 
presented in Sec.~\ref{Analogues}.
In Sec.~\ref{Systems}, these ideas are then applied to several laboratory 
systems which may allow us to recreate the fundamental effects discussed
in Sec.~\ref{Fundamental}.
Finally, Sec.~\ref{big} is devoted to the question of what can be learnt 
from studying these analogies and the outline of future directions.

\section{Fundamental Effects}\label{Fundamental}

Let us start with a brief review of the fundamental effects under 
consideration together with some of their relevant features. 
Familiarity of the reader with the basic concepts of general relativity 
and quantum field theory is assumed -- even though the main points 
should (hopefully) also become evident to all other readers. 

\subsection{Penrose Process -- Superradiance}\label{Penrose}

In a stationary (i.e., time-independent) metric, it is possible to derive 
a locally conserved energy density for the matter fields -- as expected 
from the Noether theorem. 
(Otherwise, i.e., in a genuinely time-dependent metric, there will be 
an exchange of energy between the gravitational field and the matter 
fields in general.)
For example, the conserved energy density of a massless 
(and minimally coupled) scalar field reads 
($\hbar=c=G_{\rm N}=k_{\rm B}=1$)
\bea
\label{T-special}
T^0_0=\frac{1}{2}\,g^{00}\,\dot\Phi^2-
\frac{1}{2}\,g^{ij}(\partial_i\Phi)(\partial_j\Phi)
\,.
\ea
Let us specify the second part for the equatorial plane of the 
Kerr metric \cite{Kerr} describing a rotating black hole 
\bea
\label{kerr-part}
g^{\varphi\varphi}\left(\vartheta=\frac{\pi}{2}\right)=-
\frac{{\mbox{$\displaystyle 1-\frac{2M}{r}$}}}{\Delta}
\,,
\ea
with $\Delta=r^2-2Mr+a^2$, where $M$ is the (ADM) mass of the black 
hole and $a$ measures its rotation. 
The inner and outer horizons occur at 
$r_\pm=M\pm\sqrt{M^2-a^2}$, i.e., where $\Delta=0$.
However, even outside the horizons, there is a region where 
$g^{\varphi\varphi}$ becomes negative, which is called the 
ergo-region $r_{\rm ergo}(\vartheta=\pi/2)=2M>r_\pm$.
Remembering Eq.~(\ref{T-special}), we see that the conserved 
energy density may become negative in this region. 
This fascinating observation motivates the following striking 
phenomenon:
Imagine a wave-packet stemming from spatial infinity 
(i.e., with a positive energy) which propagates into the 
ergo-region and splits up inside this region into a wave-packet 
which escapes to infinity and another part which crosses the 
(outer) horizon and is then trapped by the black hole. 
Now, if the second (trapped) part has a negative energy 
-- which is possible inside the ergo-sphere -- the remaining 
part which escapes to infinity must have an energy which is 
greater than that of the incoming wave-packet due to energy 
conservation! 
Replacing wave-packets with particles, this phenomenon is 
known as Penrose process \cite{Penrose} and for waves, 
it is called supperradiant scattering \cite{Press+Teukolsky} 
or Zel'dovich-Starobinsky effect \cite{Zeldovich+Starobinsky}. 
The explicit derivation can be done via the separation ansatz 
\bea
\label{separation-kerr}
\Phi_{\omega,\ell,m}(t,r,\vartheta,\varphi)=
e^{ -i \omega t + i m \varphi}
R_{\omega,\ell,m}(r)S_{\omega,\ell,m}(\vartheta)
\,.
\ea
Note that the applicability of this ansatz is non-trivial since
the Kerr geometry does not factorize (hence the additional $m$-dependence). 
From the conversation of the Wronskian associated to the ordinary 
differential equation for $R_{\omega,\ell,m}(r)$, one can derive the 
following relation between the reflection (i.e., field amplitude at 
spatial infinity) and transmission (at the horizon) coefficients 
\bea
1-\left|{\cal R}_{\omega,\ell,m}\right|^2=
\frac{\omega-m\Omega_{\rm h}}{\omega}
\left|{\cal T}_{\omega,\ell,m}\right|^2
\,,
\ea
where $\Omega_{\rm h}=a/(2Mr_+)$ is the angular (frame-dragging) 
velocity at the horizon.  
Consequently, for the so-called anomalous modes with 
$\omega<m\Omega_{\rm h}$, the reflection coefficient is 
larger than unity, i.e., the reflected wave is {\em stronger} 
than the incident one.
Of course, this energy gain is taken from the gravitational 
(mainly rotational) energy of the black hole, which is diminished 
by back-reaction effects (remember that the in-falling part had 
an effectively negative energy). 
Enclosing the black hole by a large ideal mirror reflecting the 
amplified modes back into the ergo-region would then generate an
blow-up instability \cite{Press+Teukolsky} (``black-hole bomb'').

\subsection{Unruh Effect}\label{Unruh}

For non-rotating black holes, the negative energies do only occur inside 
the horizon and hence cannot be exploited in such a way, if we stay on 
the classical level. 
For quantum fields, however, the local positivity of the energy density 
is a much more subtle issue and the negative-energy states inside the 
horizon {\em can} be exploited by quantum effects -- such as Hawking 
radiation. 
However, before turning to this striking phenomenon, let us study 
another quantum effect first. 

\begin{figure}[hbt]
\epsfig{file=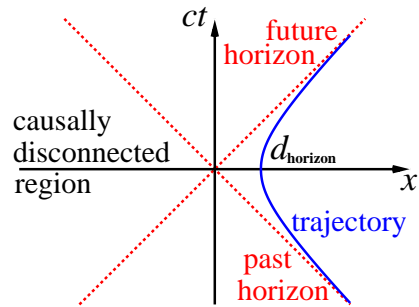,width=.3\textwidth}
\caption{Space-time trajectory with uniform acceleration and the 
associated horizons as boundaries of the Rindler wedges. 
No light ray ($45^\circ$-lines) intersecting with the trajectory 
can reach the left wedge, which is causally disconnected.}
\label{rindler}
\end{figure}

The Minkowski vacuum is (per definition) empty, i.e., free of particles,  
for all observers/detectors at rest and also (via its Lorentz invariance) 
for all inertial observers/detectors moving with a constant velocity.
Interestingly, this situations changes in the presence of 
acceleration $a$.  
Let us consider a uniformly accelerated trajectory, see Fig.~\ref{rindler}. 
Evaluating the proper (co-moving wrist-watch) time along this trajectory
we have to account for the constantly changing Lorentz boost factor  
\bea
\label{Lorentz}
\tau
=
\int dt\,\sqrt{1-\frac{\dot x^2(t)}{c^2}}
=
\frac{{\rm arsinh}(at/c)}{a/c}
\,.
\ea
Since the dynamics of a detector moving along this trajectory is 
governed by its proper time $\tau$, the (linear) response of the 
detector can be inferred from the auto-correlation function in the 
Minkowski vacuum $\ket{\rm vac}$
\bea
\label{auto-correlation}
\bra{\rm vac}\hat\Phi(\tau)\hat\Phi(\tau')\ket{\rm vac}
=
-\frac{a^2}{8\pi^2}\,\frac{1}{\cosh(a[\tau-\tau']/c)-1}
\,. 
\ea
This expression is periodic along the imaginary proper time-axis  
$\tau\to\tau+2\pi ic/a$, which motivates a thermal response of the 
detector.
Indeed, it can be shown quite generally that a uniformly accelerated 
detector experiences the inertial vacuum state as a thermal bath with 
the Unruh temperature \cite{Unruh-prd}
\bea
T_{\rm Unruh}
=
\frac{\hbar}{2\pi k_{\rm B}c}\,a
=
\frac{\hbar c}{2\pi k_{\rm B}}
\,\frac{1}{d_{\rm horizon}}
\,,
\ea
where $d_{\rm horizon}=c^2/a$ is the horizon size, i.e., the minimum 
distance between the trajectory and the causally disconnected region 
(left Rindler wedge), see Fig.~\ref{rindler}. 
The horizon experience by the accelerated observer is also the 
underlying reason for the thermal nature of the detector response:
The Minkowski vacuum $\ket{0_{\rm Min}}$ is a multi-mode squeezed 
state containing pairs of particles in either Rindler wedge 
\bea
\label{multi-mode}
\ket{0_{\rm Min}}=
\exp\left\{
\sumint{\fk{k}}
e^{-\pi\omega/a}\,
\hat a^\dagger_{\rm R,\,left}(\omega)\,
\hat a^\dagger_{\rm R,\,right}(\omega)
\right\}
\ket{0_{\rm FR}}
\,,
\ea
where $\omega=\omega(\f{k})>0$ is understood.
Here $\ket{0_{\rm FR}}$ denotes the Fulling-Rindler vacuum, 
which is the local ground state of the accelerated detectors. 
Since the accelerated observer cannot see the left Rindler wedge,
we may trace over the degrees of freedom of the left wedge, 
which turns the pure state $\ket{0_{\rm Min}}$ into a mixed state --
which precisely corresponds to a thermal density matrix 
(thermo-field formalism \cite{Israel}).

\subsection{Hawking Radiation}\label{Hawking}

The Unruh effect described above is strongly related to 
Hawking radiation \cite{Hawking} via the principle of equivalence. 
In fact, Bill Unruh \cite{Unruh-prd} discovered it when trying to 
understand black hole evaporation. 
Imagine an imprudent astronaut freely falling towards (and finally through) 
the horizon.
Locally, this freely falling astronaut is equivalent to an inertial 
observer.  
Another astronaut staying at a fixed distance to the horizon, however,
feels the gravitational pull and thus corresponds to a uniformly 
accelerated observer.
Now he/she can compare measurements results with his/her unfortunate 
colleague. 
Assuming that the freely falling astronaut does not experience anything 
special when passing the horizon, i.e., sees a quantum state locally 
equivalent to the Minkowski vacuum (as one would expect), the results 
of the previous section indicate that the stationary observer should 
see a thermal spectrum. 
Indeed, assuming a regular quantum state, the black hole emits thermal 
radiation with the Hawking temperature 
\bea
\label{Hawking-original}
T_{\rm Hawking}
=
\frac{1}{8\pi M}
\frac{\hbar\,c^3}{G_{\rm N}k_{\rm B}}
=
\frac{\hbar}{2\pi k_{\rm B}c}\,a_{\rm surface\;gravity}
\,,
\ea
where the surface gravity measures the gravitational acceleration 
at the horizon, see the following table:  

\bigskip
\begin{tabular}{cc}
\hline
\noalign{\smallskip}
Unruh effect & Hawking effect \\
\noalign{\smallskip}
\hline
\noalign{\smallskip}
flat space-time & black hole \\
\noalign{\smallskip}
\hline
\noalign{\smallskip}
accelerated observer & observer at a fixed distance \\
\noalign{\smallskip}
\hline
\noalign{\smallskip}
inertial observer & freely falling observer \\
\noalign{\smallskip}
\hline
\end{tabular}
\bigskip

However, in spite of the analogy sketched above, there are also crucial 
differences between the Unruh effect and Hawking radiation:
While the Unruh set-up is invariant under time-reversal
(the acceleration remains unchanged after $t\to-t$), 
the Hawking scenario breaks this symmetry -- 
one can only fall into the black hole, but never come out
(see Sec.~\ref{Eardley} below). 
As a consequence, the Unruh effect is associated to a truly thermal bath
containing an equal amount of particles in both directions and thus no 
net flux.
In contrast, Hawking radiation corresponds to a real energy out-flux 
(black hole evaporation) and does not generate particles propagating 
from spatial infinity into the black hole. 
Furthermore, propagating the particles of the Hawking radiation back in 
time and thereby undoing the gravitational red-shift, we find that they
originate from modes with shorter and shorter wavelengths in the past.  
Within the framework of quantum fields in classical curved space-times, 
there is no limit to this time-reversed squeezing process and thus these
modes originate from an energy region (perhaps the Planck scale) where 
this semi-classical treatment breaks down and effects of quantum gravity
become important. 
Ergo, Hawking's prediction is based on the extrapolation of a theory 
(quantum fields in classical curved space-times) into an energy region 
where it is expected to break down.
This observation poses the question of whether the Hawking effect 
survives after incorporating the effects of quantum gravity at high
energies (trans-Planckian problem).

Let us study this question by implementing a change of the dispersion 
relation $\omega=ck\to\omega(k)$ as a toy model \cite{Jacobson}
for high-energy deviations from classical general relativity. 
Using the Eddington-Finkelstein coordinates
\bea
ds^2=\left(1-\frac{2M}{r}\right)dv^2-2dv\,dr
\,,
\ea
the wave equation of a massless scalar field reads 
\bea
\left(2\partial_v\partial_r+
\partial_r\left[1-\frac{2M}{r}+f(-\partial_r^2)\right]\partial_r
\right)\Phi=0
\,,
\ea
where we have added a modification $f(k^2)$ of the dispersion relation. 
After a Fourier-Laplace transformation, we get the following integral 
equation
\bea
\left(2\omega-k\left[1+2iM\partial_k^{-1}+f(k^2)\right]\right)
\phi_\omega(k)=0 
\,,
\ea
which can be solved via separation of variables
\bea
\label{inverse}
\phi_\omega(r)=\int\frac{dk}{g(k)}\,
\exp\left\{ikr-2iM\int\frac{dk'}{g(k')}\right\}
\,,
\ea
with the spectral function $g(k)=1+f(k^2)-2\omega/k$.
The integrand becomes singular at $k_*$ where $g(k_*)=0$, 
which is just the solution of the dispersion relation far 
away from the black hole $r\to\infty$. 
Assuming that $f(k^2)$ is an analytic function, we may calculate the 
integral by deforming the integration contour into the complex plane
\cite{universality}. 
Since $M$ is supposed to be very large (compared to the Planck mass 
and $\omega$ etc.), we may use the saddle-point method.
We get two saddle points at large and real wavenumbers of opposite 
sign $k_\pm$, which are (as one would expect) solutions of the 
dispersion relation for finite $r$. 
These are the in-modes $\phi_{\rm in}^\pm(\omega)$ and we shall assume 
that these short-wavelength modes are in their ground state
$\hat a_{\rm in}(\omega)\ket{\rm in}=0$. 
Finally, for closing the integration contour, we have to circumvent 
the branch cut originating from the singularity at $k_*$, which just 
yields the outgoing long-wavelength Hawking modes $\phi_{\rm out}$ 
at $r\to\infty$. 
Now we may calculate the particle content in these out-modes 
$\phi_{\rm out}$ by connecting them to the in-modes 
$\phi_{\rm out}(\omega)\leftrightarrow\alpha_\omega
\phi_{\rm in}^+(\omega)+\beta_\omega\phi_{\rm in}^-(\omega)$.  
The Bogoliubov coefficient can be obtained by comparing the 
saddle-point contributions at $k_+$ and $k_-$ and determine 
the particle number via 
$\hat a_{\rm out}(\omega)\leftrightarrow\alpha_\omega
\hat a_{\rm in}(\omega)+\beta_\omega\hat a_{\rm in}^\dagger(\omega)$ 
which implies 
$\bra{\rm in}\hat a_{\rm out}(\omega)\hat a_{\rm out}^\dagger(\omega)
\ket{\rm in}=|\beta_\omega^2|$.
Comparison with the Bose-Einstein distribution yields the effective 
Hawking temperature for each frequency \cite{origin}
\bea
\label{temperature}
T_{\rm Hawking}(\omega)
=
\frac{v_{\rm gr}(k_*)v_{\rm ph}(k_*)}{8\pi M}
\,,
\ea
where $k_*$ denotes the solution of the dispersion relation for a given 
$\omega$ far away from the black hole $r\to\infty$ while $v_{\rm gr}$
and $v_{\rm ph}$ are the group and phase velocities at that frequency. 
If the dispersion relation becomes linear $\omega=ck$ at small $k$ or 
approaches that of a massive particle $\omega^2=m^2c^4+c^2k^2$, we 
reproduce Eq.~(\ref{Hawking-original}). 
However, if the dispersion relation $\omega(k)$ rises too fast 
(i.e., faster than quadratic), we would get an increasing amount of 
radiation at high energies (``UV-catastrophe'') and the black hole 
would evaporate quickly.
Such a case should be excluded in view of our observational evidence 
of black holes with macroscopic life-times. 
 
\subsection{Eardley Instability and Black-hole Laser}\label{Eardley}

After altering the dispersion relation, the Hawking particles do no longer
originate from modes with arbitrarily short wavelengths squeezed against 
the horizon.
Depending on the curvature of the dispersion relation, the short-wavelength
modes are faster (superluminal case) or slower (subluminal case) than the
speed of light $c$. 
Hence the in-modes with $k_\pm$ approach the horizon from the inside 
(superluminal case) or the outside (subluminal case).
During that process, they are stretched due to the gravitational 
red-shift (tidal forces) and their wavelength grows, i.e., their 
group and phase velocity approaches $c$. 
Finally, they are ripped apart by the horizon and one part 
(the Hawking particle) escapes to infinity while the remaining part  
(the in-falling partner particle) is trapped. 
This the in-falling partner particle has a negative energy 
(cf.~Sec.~\ref{Penrose}) and thereby compensates the energy emitted 
by the Hawking radiation. 

As mentioned above, this process is not symmetric under time-reversal.
Now let us consider the time-reversed situation:
Time-reversal turns a black hole (where nothing can escape) into a 
white hole (which nothing can penetrate). 
White holes are unstable solutions of the Einstein equations -- 
while they still exert the gravitational attraction to everything, 
they do not allow anything to pass the white-hole horizon.
Naturally, this leads to a pile-up at the horizon (until the dispersion 
relation changes or non-linear effects set in).
Repeating the same analysis as before, we find that an in-going 
low-wavenumber vacuum state is transformed to outgoing high-wavenumber 
excitations.  
This means that white holes would decay very rapidly or turn into 
black holes (Eardley instability \cite{Eardley}).

Another interesting effect may occur when a black-hole horizon is 
combined with a white-hole horizon, which effectively happens in the 
Reissner-Nordstr\"om metric describing a charged black hole. 
In this case, the negative-energy wave packet may bounce back and forth 
between the two horizons where it emits a Hawking wave each time it 
hits the black-hole horizon and thereby constantly grows
(similar to the ``black-hole bomb'' in Sec~\ref{Penrose}). 
This instability is called a black-hole laser \cite{laser}. 

\subsection{Cosmological Particle Creation}\label{Cosmological}

The phenomenon of particle creation is not restricted to the gravitational
field around black (or white) holes, but may also occur in an expanding 
(or contracting) universe. 
Let us specify the action of a scalar field 
\bea
\label{action}
{\cal A}=\frac12
\int d^4x\,\sqrt{|{\mathfrak g}|}
\left[
(\partial_\mu\Phi){\mathfrak g}^{\mu\nu}(\partial_\nu\Phi)
-\zeta{\mathfrak R}\Phi^2
\right]
\,,
\ea
for the Friedmann-Robertson-Walker metric  
\bea
\label{metric}
ds^2={\mathfrak a}^6(t)dt^2-{\mathfrak a}^2(t)d\f{r}^2
\,,
\ea
where ${\mathfrak a}(t)$ is the time-dependent scale factor of the 
Universe and ${\mathfrak R}$ the Ricci curvature scalar
($\zeta$ is the dimensionless curvature coupling factor, e.g., 
$\zeta=1/6$ for conformal coupling in 3+1 dimensions and $\zeta=0$ 
for minimal coupling).
The wave equation reads 
\bea
\label{normal-mode}
\ddot\Phi_{\fk{k}}+
\left[
{\mathfrak a}^4(t)\f{k}^2+\zeta{\mathfrak a}^6(t){\mathfrak R}(t)
\right]
\Phi_{\fk{k}}=0
\,.
\ea
Accordingly, each $\f{k}$-mode behaves as a harmonic oscillator with 
a time-dependent potential $\Omega^2_{\fk{k}}(t)$ given by 
$\Omega^2_{\fk{k}}(t)=
{\mathfrak a}^4(t)\f{k}^2+\zeta{\mathfrak a}^6(t){\mathfrak R}(t)$.
Now, if the temporal variation of $\Omega_{\fk{k}}(t)$ 
is much slower than the internal frequency 
($\dot\Omega_{\fk{k}}\ll\Omega_{\fk{k}}^2$ and 
$\ddot\Omega_{\fk{k}}\ll\Omega_{\fk{k}}^3$ etc.), 
the quantum state of the harmonic oscillator will stay near the 
ground state due to the adiabatic theorem. 
However, if the external time-dependence $\Omega_{\fk{k}}(t)$ 
generated by the cosmic expansion is too fast (i.e., non-adiabatic), 
the ground-state wave-function cannot adapt to this change anymore 
and thus the evolution will transform the initial ground state into 
an excited state in general (cosmological particle creation). 
Since Eq.~(\ref{normal-mode}) is linear, the excited state is a 
squeezed state 
\bea
\label{squeezed}
\ket{\psi(t\uparrow\infty)}
=
\exp\left\{\sum\limits_{\fk{k}}
\left(
\xi_{k}\hat a_{\fk{k}}\hat a_{-\fk{k}}
-{\rm h.c.}
\right)
\right\}\ket{0}
\,,
\ea
i.e., the particle are created in pairs of opposite wavenumbers 
due to momentum conservation. 

Basically the same mechanism is (according to our standard model of 
cosmology) responsible for the generation of the seeds for structure 
formation out of the quantum fluctuations of the inflaton field. 
This process happened during inflation and at the end of this epoch,
the amplified (squeezed) quantum fluctuations were transformed into 
density/temperature variations via the decay of the inflaton field. 
Traces of these primordial density/temperature variations can still 
be observed today in the anisotropies of the cosmic microwave 
background radiation. 

\subsection{Gibbons-Hawking Effect}\label{Gibbons-Hawking}

In contrast to cosmological particle creation discussed above, 
where the produced particles can be detected {\em after} the expansion, 
the Gibbons-Hawking effect \cite{Gibbons+Hawking} concerns the response 
of a detector {\em during} the expansion and is more similar to the 
Unruh effect. 
In order to point out the difference, let us rewrite the 
Friedmann-Robertson-Walker metric (\ref{metric}) in terms of the 
conformal time $\eta$
\bea
\label{conformal}
ds^2={\mathfrak a}^2(\eta)[d\eta^2-d\f{r}^2]
\,,
\ea
and consider a conformally invariant field, e.g., massless scalar field 
in 1+1 dimensions or the electromagnetic field in 3+1 dimensions
\bea
\label{em-action}
{\cal A}_{\rm matter}=-\frac14\int d^4x\,\sqrt{-g}\,
F_{\mu\nu}F^{\mu\nu}
\,,
\ea
where $F_{\mu\nu}$ is the field strength tensor 
\bea
\label{em-field}
F_{\mu\nu}=
\nabla_\mu A_\nu-\nabla_\nu A_\mu=
\partial_\mu A_\nu-\partial_\nu A_\mu
\,.
\ea
The last equality exploits the symmetry of the Christoffel symbols 
$\Gamma^\rho_{\mu\nu}=\Gamma^\rho_{\nu\mu}$.
Inserting the metric (\ref{conformal}) into the action (\ref{em-action}),
we see that all factors of ${\mathfrak a}(\eta)$ cancel and thus the 
field modes in terms of $\eta$ are just undisturbed plane waves.
Consequently, there is no cosmological particle creation in the sense 
of Sec.~\ref{Cosmological} (conformal vacuum). 
However, this does not mean that a photon detector in such an expanding 
Universe would not click.
In contrast to the photon modes, the detector dynamics is not labeled 
by the conformal time $\eta$ but by the proper time 
$d\tau^2={\mathfrak a}^2(\eta)d\eta^2$.
Similar to Sec.~\ref{Unruh}, the two-point correlation function 
$\propto1/(\eta-\eta')^2$ along the detector's trajectory $\f{r}=0$ 
changes after transformation to the proper time $\tau$ and hence 
the detector will detect excitations in general. 
For example, in the de~Sitter Universe 
${\mathfrak a}(\tau)=\exp\{{\mathfrak H}\tau\}$ possessing a constant 
Hubble parameter ${\mathfrak H}=\dot{\mathfrak a}/{\mathfrak a}$, 
the auto-correlation function of the detector behaves as 
$\eta\eta'/(\eta-\eta')^2$, where the scale factors 
${\mathfrak a}(\eta){\mathfrak a}(\eta')=1/\eta\eta'$ stem from the 
coupling to the physical degrees of freedom of the detector. 
In terms of the proper time $\tau\propto\ln\eta$, this translates 
into $\propto1/\sinh^2({\mathfrak H}[\tau-\tau'])$, which is analogous to 
Eq.~(\ref{auto-correlation}). 
Again, we observe a periodicity in imaginary $\tau$-direction 
and hence expect a behavior analogous to Unruh effect.
Indeed, it can be shown \cite{Gibbons+Hawking} that the proper-time 
detector experiences the conformal vacuum as a thermal bath with a 
temperature 
\bea
T_{\rm Gibbons-Hawking}
=
\frac{\hbar}{2\pi k_{\rm B}}\,{\mathfrak H}
\,.
\ea
%

\subsection{Schwinger Mechanism}

It is instructive to identify common features of the various effects 
considered above.
Apart from first example in Sec.~\ref{Penrose} 
(and, in some sense, the one in  Sec.~\ref{Eardley}),
they are pure quantum effects which do not occur on the classical level. 
E.g., the classical vacuum $\Phi\equiv0$ remains empty $\Phi\equiv0$
for all non-inertial observers.
Furthermore, the effects in Sec.~\ref{Penrose} 
(and \ref{Eardley}, of course) do also have their quantum analogues: 
For quantized fields, it is not necessary to send a wave-packet into
the rotating black hole in order to obtain superradiance, the in-going 
vacuum fluctuations are sufficient to create outgoing radiation. 
In all of these quantum effects, the in-going ``virtual'' quantum vacuum 
fluctuations are converted into ``real'' 
(w.r.t.~the detector under consideration) particle pairs by the 
external influence (e.g., the gravitational field).
These pairs are correlated (even entangled) and can be described by a 
(multi-mode) squeezed state such as in Eq.~(\ref{multi-mode}). 
For cosmological particle creation, the two particles have opposite 
momenta, cf.~Eq.~(\ref{squeezed}), whereas, for the Unruh effect, they 
live in distinct Rindler wedges, i.e., on different sides of the horizon. 
The same is true for the Gibbons-Hawking effect and the black-hole case
in Sec.~\ref{Hawking} (as well as Sec.~\ref{Penrose}).
In the latter situation each pair consists of an outgoing (Hawking) 
particle and its in-falling partner (with negative energy). 

In addition, all of our examples are relativistic quantum effects. 
Thus, for typical laboratory parameters, these effects are suppressed 
due to large value of $c$ and the smallness of $\hbar$ 
(and partly $G_N$).
Finally, for stationary scenarios, the effects mentioned above can be 
understood in terms of the population of negative energy states
(which liberates the energy necessary for creating a ``real'' particle). 

From this point of view, there is yet another effect, which also fits 
into this scheme: the Schwinger mechanism 
\cite{Dirac,Klein,Sauter,Heisenberg+Euler,Schwinger}. 
In analogy to the gravitational field, which rips apart the wave packet 
near the horizon, a strong enough electric field may also pull an 
electron-positron pair out of the vacuum.
This pair creation induced by the strong electric field is the main 
mechanism for neutralization of charged black holes (i.e., it is stronger 
than Hawking radiation for significantly charged black holes).
However, it is not tied to black holes and might also occur in the 
laboratory.
Let us consider a constant electric field $\f{E}=E\f{e}_x$ 
for simplicity.
In this case, the dispersion relation of the electrons/positrons reads 
\bea
\label{disp-E}
(\omega\pm qEx)^2=m^2+k^2
\,.
\ea
As a result, a classical electron trajectory with a given positive 
energy $\omega$ is incident from the right and then totally reflected 
(at the classical turning point). 
Positrons can be pictured as holes ion the Dirac sea \cite{Dirac} 
(corresponding to negative-energy states) and hence obey the opposite
behavior, i.e., they are incident from the 
left and totally reflected back to the left. 
The two sets of classical trajectories are separated by the gap 
between the positive and negative energy states.
For $E=0$, this gap is given by 
$\omega_{\rm pos}\geq+m$ and $\omega_{\rm neg}\leq-m$. 
In the presence of an electric field $E>0$, however, the gap is tilted 
and becomes $x$-dependent -- which just correspond to the two classical 
turning points.
In a quantum description, however, an electron from the Dirac sea may
tunnel through the gap and become a real particle -- leaving behind a 
hole, i.e., a positron.
If the field $E$ is not too strong, we may estimate the tunneling 
probability via the WKB approximation. 
For $\omega=k_y=k_z=0$, we get the eikonal of the exponential tail in 
the classically forbidden region in analogy to tunneling trough a 
barrier from Eq.~(\ref{disp-E})
\bea
\label{exponent}
\int\limits_{-m/(qE)}^{+m/(qE)} dx\,\sqrt{m^2-(qEx)^2}=
\pi\,\frac{m^2}{2qE}
\,.
\ea
Hence the tunneling probability is exponentially suppressed 
$\propto\exp\{-\pi E_S/E\}$, where $E_S=m^2/q$ is called the 
Schwinger limit.
At that ultra-high field strength, the work done separating the 
electron-positron pair over a Compton wavelength is of the order 
of the binding energy $2m$ and higher-order processes become 
important, i.e., the above estimate fails. 

\section{Laboratory Analogues}\label{Analogues}

The effects sketched in the previous sections are fundamental predictions 
of quantum field theory in non-trivial (e.g., gravitational) background
configurations.  
Testing these predictions would be extremely interesting -- but, 
unfortunately, none of these effects has been directly observed yet.
(So far, there are only indirect signatures, such as the anisotropies 
of the cosmic microwave background radiation mentioned in 
Sec.~\ref{Cosmological}.)
This lack of experimental evidence is mainly due to the suppression of 
these effects by the smallness of $\hbar$ and $1/c$ and the resulting 
difficulty of creating strong enough fields or accelerations etc.  
One way of overcoming this obstacles are analogues using sound waves 
(or other quasi-particles) instead of photons (or electrons/positrons) 
and thereby effectively replacing the speed of light $c$ by the speed 
of sound $c_{\rm s}\lll c$.
Of course, it would also be very interesting to find scenarios with an 
effectively increased $\hbar$ in addition, but this option will not be 
discussed here. 

\subsection{The Acoustic Analogy}

Analogies between laboratory systems and gravitational fields have been 
known for a long time (e.g., the Gordon metric \cite{Gordon}), but it was 
probably Bill Unruh \cite{Unruh-prl} who first realized that one could 
exploit these similarities for studying fundamental effects such as 
Hawking radiation. 
He considered sound waves $\delta\vau=\na\Phi$ in irrotational fluids 
$\na\times\vau=0$, which obey the wave equation 
\bea
\left(\frac{\partial}{\partial t}+\na\cdot\vau_0\right)
\frac{\varrho_0}{c^2_{\rm s}}
\left(\frac{\partial}{\partial t}+\vau_0\cdot\na\right)
\Phi
=
\na\cdot(\varrho_0\na\Phi)
\,,
\ea
where $\varrho_0$ and $\vau_0$ denote the density and the velocity 
of the background flow and $c_{\rm s}$ is the speed of sound. 
The analogy to gravity is based on the striking observation that 
this wave equation is quantitatively equivalent to the 
d'Alembertian in a curved space-time 
\bea
\Box_{\rm eff}\Phi
=
\frac{1}{\sqrt{-g_{\rm eff}}}\partial_\mu
\left(\sqrt{-g_{\rm eff}}\,g^{\mu\nu}_{\rm eff}\partial_\nu\Phi\right)
=0
\,,
\ea
if the effective (acoustic) metric $g^{\mu\nu}_{\rm eff}$
is chosen in a specific way, 
which is similar to the Painlev{\'e}-Gullstrand-Lema{\^\i}tre form 
\cite{PGL} known from gravity 
\bea
\label{PGL}
g^{\mu\nu}_{\rm eff}
=
\frac{1}{\varrho_0c_{\rm s}}
\left(
\begin{array}{cc}
1 & \vau_0 \\
\vau_0 & \vau_0\otimes\vau_0 - c^2_{\rm s}\f{1}
\end{array}
\right)
\,.
\ea
As a result, phonons propagating in this flowing fluid behave precisely 
in the same way as a scalar (quantum) field in a curved space-time 
described by the above metric.
Even though the form (\ref{PGL}) is not flexible enough to reproduce 
{\em all} possible gravitational fields (e.g., it cannot simulate the 
full Kerr geometry), it is sufficiently general to model rotating and 
non-rotating black or white holes 
(cf.~Secs.~\ref{Penrose}, \ref{Hawking}, and \ref{Eardley})
and an expanding or contracting universe (cf.~Sec.~\ref{Cosmological}).
The Unruh and the Gibbons-Hawking effect discussed in 
Secs.~\ref{Unruh} and \ref{Gibbons-Hawking}, respectively, 
require a more detailed consideration of the phonon detector. 
Recreating the Schwinger mechanism necessitates the analogue of a 
Dirac sea and thus fermionic quasi-particles as well as the coupling 
to an electric field analogue.
(An experiment along these lines has been done in graphene \cite{graphene}. 
However, the quasi-particle spectrum does not exhibit a mass gap in 
this system and hence the analogy to the Schwinger mechanism is limited.)

\subsection{Further Analogues}\label{Further}

The analogy between condensate matter and gravity is not restricted to sound 
waves, it may apply to other quasi-particles as well.
If the quasi-particles are Goldstone-modes (i.e., gap-less) and can be 
described by a single scalar field $\Phi$, the most general linearized 
low-energy effective action reads \cite{Barcelo}
\bea
\label{general}
{\mathcal L}_{\rm eff}
=
\frac12(\partial_\mu\Phi)(\partial_\nu\Phi)G^{\mu\nu} 
+\ord(\Phi^3)
+\ord(\partial^3)
\,,
\ea
where $G^{\mu\nu}$ is governed by the underlying condensed-matter 
system and may depend on space and time in general.
Identifying $G^{\mu\nu}\to g^{\mu\nu}_{\rm eff}\sqrt{|g_{\rm eff}|}$, 
we again find a qualitative analogy to gravity. 
Apart from phonons, examples for such quasi-particles are surface 
waves (ripplons) on top of flowing liquids \cite{ripplon} or photons 
with a fixed polarization in wave guides \cite{wave-guide} or optical 
fibers \cite{Philbin}. 
Under additional assumptions, the analogy can also be extended to 
non-scalar quasi-particles.
One example are photons in a moving dielectric medium with an arbitrary 
four-velocity $u^\mu$ and a constant (scalar) permittivity $\varepsilon$, 
whose behavior is analogous to a curved space-time with the Gordon 
metric \cite{Gordon}
\bea
\label{gordon-o}
g^{\mu\nu}_{\rm eff}=g^{\mu\nu}_{\rm Min}
+(\varepsilon-1)\,u^\mu\,u^\nu
\,.
\ea
It is even possible to model the 1+1 dimensional Dirac equation 
(i.e., spin-1/2 particles) within a slow-light set-up \cite{Slow}, 
but these excitations do still obey bosonic commutation relations,
i.e., there is no Dirac sea.  

\subsection{Analogue Gravity}

Let us apply the concepts and tools from general relativity to the 
acoustic metric (\ref{PGL}); most of the other effective metrics  
are very similar.
Comparing Eqs.~(\ref{PGL}) and (\ref{T-special}), we see that an 
ergo-region occurs for supersonic flow velocities $\vau_0^2>c^2_{\rm s}$. 
A horizon, on the other hand, corresponds to a closed surface at which 
the {\em normal} flow velocity exceeds the speed of sound 
$v_\perp=c_{\rm s}$.
For a radial or effectively one-dimensional flow, the two points 
coincide, but for a maelstrom-type flow (which can still be locally 
irrotational), there is a finite ergo-region outside the horizon.
Since the effective energy of the phonons may become negative in this 
region, we may obtain superradiance effects. 

\begin{figure}[hbt]
\epsfig{file=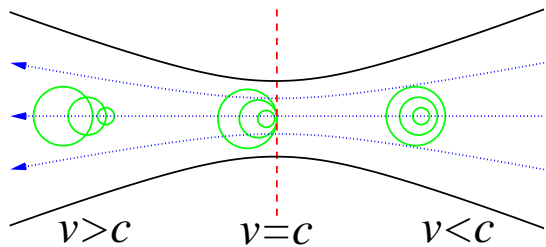,width=.4\textwidth}
\caption{Sketch of a de~Laval nozzle as an analogue for a black hole.
The solid (black) lines denote the walls of the nozzle and the dotted 
(blue) curves are the streamlines of the flow. 
At the entrance of the nozzle (right), the flow is sub-sonic $v<c$.
The flow velocity exceeds the speed of sound at the narrowest point 
and exits the nozzle (left) with a super-sonic velocity. 
The phonons are depicted by (green) circles.
In the sub-sonic region (right), they may propagate in both directions 
-- but in the supersonic part (left), they are swept away and cannot 
escape anymore. 
Hence the border (red dashed line) between the two regions is analogous 
to a black hole horizon with the right half of the nozzle corresponding 
to the outside of the black hole and left half modelling the black hole 
interior.}
\label{laval}
\end{figure}

If the flow is accelerated (following a streamline) form subsonic to 
supersonic velocities (such as in a de~Laval nozzle, cf.Fig.~\ref{laval}), 
we obtain the analogue of a black-hole horizon, which traps all sound 
waves.
The time-reverse, i.e., a flow decelerated from supersonic back to 
subsonic speed, would correspond to a white-hole horizon, which expels 
all phonons. 
As one would expect from the discussion in Sec.~\ref{Eardley}, such a 
flow profile is typically plagued with instabilities 
(towards the generation of shock waves etc.). 

\begin{figure}[hbt]
\epsfig{file=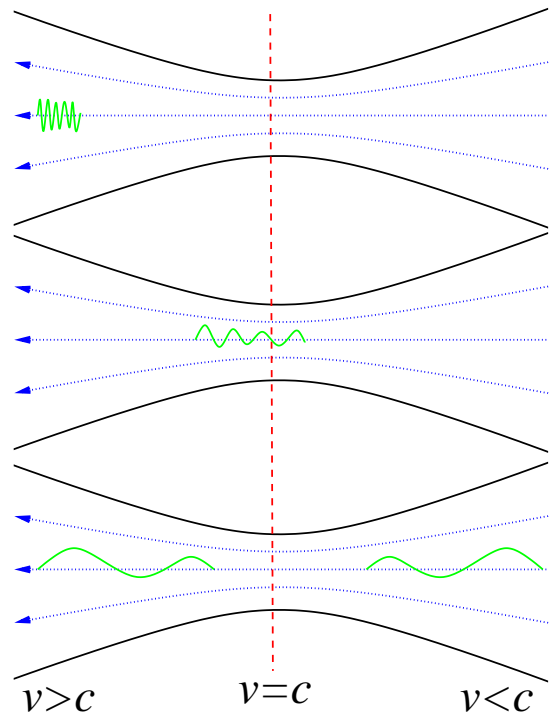,width=.4\textwidth}
\caption{Evolution of the wave-packet generating the Hawking radiation in 
a de~Laval nozzle for a super-sonic dispersion relation.
Initially, the short-wavelength wave-packet 
($k_\pm$ in Sec.~\ref{Hawking}) with a large group velocity
$d\omega/dk>c$ is propagating upstream and approaches the horizon from 
the inside (top panel).
During that process, the flow inhomogeneity stretches the wave-packet
(middle panel) in analogy to the gravitational red-shift and thereby 
lowers its group velocity until $d\omega/dk\approx c$. 
Finally, the wave-packet is ripped apart by the horizon into two parts
(bottom panel), the outgoing Hawking radiation 
($k_*$ in Sec.~\ref{Hawking}) and its in-falling 
partner particle (both with long wavelengths).}
\label{laval-stages}
\end{figure}

Generating the analogue of a black hole horizon in a de~Laval nozzle,
for example, the same arguments as in Hawking's original derivation 
\cite{Hawking} would imply a thermal phonon flux moving upstream. 
Even if we send in the fluid at zero temperature, the in-going quantum 
fluctuations of the phonon modes will be ripped apart by the horizon 
and generate excitations in the form of phonon pairs. 
(One phonon escapes to the entrance of the nozzle and the other one 
is swept away toward its exit.)
Inserting the acoustic metric (\ref{PGL}), the surface gravity is 
determined by the velocity gradient at the horizon and thus the 
Hawking temperature reads 
\bea
\label{gradient}
T_{\rm Hawking}
=
\frac{\hbar}{2\pi\,k_{\rm B}}\,
\left|
\frac{\partial}{\partial r}\left(v_0-c_{\rm s}\right)
\right|_{v_0=c_{\rm s}}
\,.
\ea
Inserting typical values for potential condensed matter analogues, 
we obtain relatively low temperatures -- which are, however, still 
orders of magnitude larger than for astronomical black holes. 
Of course, the trans-Planckian problem does also play a role here,
because the Hawking phonons originate from modes with very short 
wavelengths, where fluid dynamics breaks down.
In contrast to (quantum) gravity, we do know (at least in principle) 
the correct microscopic theory for fluids -- and thus we may address 
the trans-Planckian problem.
If a change of the dispersion relation describes the relevant deviation
from fluid dynamics at short length scales, we may employ the methods
sketched in Sec.~\ref{Hawking}, cf.Fig.~\ref{laval-stages}). 
Other effects (such as non-linear interactions), however, are much 
less understood. 

Finally, the metric (\ref{PGL}) can also model an expanding or 
contracting universe.
There are basically two possibilities (which can be combined):
an expansion or contraction of the fluid and variations in the 
local speed of sound. 

\section{Laboratory Systems}\label{Systems}

Now, after having briefly discussed the underlying analogy, we are in the 
position to study some explicit condensed-matter examples.
The following selection is not exhaustive, but aimed at providing a rough
overview and indicating possible future directions. 

\subsection{Trapped Ions}\label{Ions}

Let us start with a system where the technology for cooling to the ground 
state and detecting single particles is most advanced -- a chain of ions 
in a trap. 
Assuming that the radial confinement is very tight, it is sufficient to 
consider the positions $q_i$ of the ions in axial direction.
In the presence of a time-dependent harmonic axial trap potential
$\omega_{\rm ax}^2$, the equations of motion read \cite{ion-trap}
\bea
\label{eom-full}
\ddot q_i(t)+\omega_{\rm ax}^2(t)q_i(t)
=
\gamma\sum\limits_{j \neq i}
\frac{{\rm sign}(i-j)}{[q_i(t)-q_j(t)]^2}
\,,
\ea
where $\gamma$ denotes the strength of the Coulomb repulsion. 
The classical equations of motion can be solved in terms of a 
single scale parameter given by 
\bea
\label{scaling}
\ddot b(t)+\omega_{\rm ax}^2(t)b(t)
=
\frac{\omega_{\rm ax}^2(0)}{b^2(t)}
\,.
\ea
In order to obtain the phonon modes, we split the full quantum position 
operator $\hat q_i(t)$ into the classical solution $b(t)q_i^0$ which is 
fully determined by the initial static position $q_i^0$ plus small 
(linearized) quantum fluctuations 
$\hat q_i(t)=b(t)q_i^0+\delta\hat q_i(t)$. 
After linearization and normal-mode decomposition, we obtain the 
phonon modes 
\bea
\label{kappa}
\left(
\frac{\partial^2}{\partial t^2}+\omega_{\rm ax}^2(t)
+\frac{\omega_\kappa^2}{b^3(t)}
\right)
\delta\hat q_\kappa
=0
\,,
\ea
where $\omega_\kappa^2$ are the time-independent eigenfrequencies 
of the phonon modes labelled by $\kappa$. 
Identifying 
$\kappa\leftrightarrow\f{k}$, 
$\omega_\kappa^2\leftrightarrow\f{k}^2$, 
$b(t)\leftrightarrow{\mathfrak a}(t)$, 
$\omega_{\rm ax}^2(t)\leftrightarrow{\mathfrak R}(t)$,
we obtain a striking similarity to Eq.~(\ref{normal-mode}).
Moreover, both, ${\mathfrak R}(t)$ and $\omega_{\rm ax}^2(t)$, 
are related to the second time-derivatives of the corresponding
scale factors ${\mathfrak a}(t)$ and $b(t)$.
As a result of the analogy between Eqs.~(\ref{kappa}) and 
(\ref{normal-mode}), we obtain the analogue of cosmological 
particle creation in an ion trap with a time-dependent axial 
trapping strength $\omega_{\rm ax}^2(t)$.
Accordingly, the initial phonon creation and annihilation operators 
$\hat a_\kappa^\dagger(0)$ and $\hat a_\kappa(0)$ are related to the 
final phonon operators via the Bogoliubov transformation 
\bea
\label{Bogolubov}
\hat a_\kappa(t\uparrow\infty)
=
\alpha_\kappa\hat a_\kappa(0)
+\beta_\kappa\hat a_\kappa^\dagger(0)
\,,
\ea
which exactly corresponds to the squeezed state in Eq.~(\ref{squeezed}) 
via 
$|\alpha_\kappa|\leftrightarrow\cosh(|\xi_{\fk{k}}|)$
and 
$|\beta_\kappa|\leftrightarrow\sinh(|\xi_{\fk{k}}|)$.

\begin{figure}[hbt]
\epsfig{file=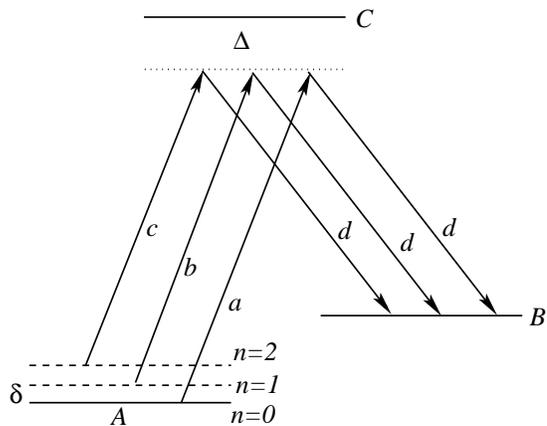,width=.4\textwidth}
\caption{Sketch (not to scale) of the relevant electronic 
(vertical solid lines) and motional (vertical dashed lines) levels 
and the associated transitions (arrows).}
\label{lambda}
\end{figure}

The created phonon pairs can be detected via suitable red-sideband Raman 
transitions, see Fig.~\ref{lambda}.
Initially, the ions are in the electronic ($A$) and motional ground state 
($n=0$). 
Then applying a Raman $\pi$-pulse ($a+d$) would transfer the full 
population into the meta-stable state $B$ without actually populating 
the excited state $C$ due to the large detuning $\Delta$.
If we apply an additional detuning $n\delta$ to one of the Laser beams 
only, i.e., a first ($n=1$) or second ($n=2$) side-band transition 
$b+d$ or $c+d$, respectively, we may selectively transfer population 
from the excited motional levels $n=1$ or $n=2$ only to the state $B$. 
Since the occupation of that meta-stable state $B$ could be measured 
by suitable detected by resonance fluorescence methods, we may infer 
the initial motional level. 
It is even possible to distinguish two-phonon states $n=2$ from 
one-particle excitations $n=1$, which could be used to discriminate
the quantum effect (squeezed state with phonon pairs) from classical
effects such as heating \cite{ion-trap}. 

In contrast to the above analogue to cosmological particle creation 
in an ion trap, where the produced phonons are measured {\em after} 
varying the trap frequency $\omega_{\rm ax}^2(t)$, there has been 
another proposal for detecting phonons {\em during} the expansion
of the ion cloud \cite{Alsing}. 
However, the latter proposal poses several technical and conceptual 
problems: 
Since the phonon modes are not conformally invariant, one would obtain 
some sort of combination of the Gibbons-Hawking effect and cosmological 
particle creation, cf.~Eq.~(\ref{Bogolubov}), which has not been 
taken into account sufficiently. 
Furthermore, resolving the phonon energies $\delta$ requires a 
sufficiently long measurement time 
$\Delta t=\ord(\hbar/\Delta E)=\ord(\hbar/\delta)$. 
Sustaining an exponential expansion during such a long time is of 
course very difficult.
Finally, the ion dynamics (\ref{scaling}) was not included adequately. 

In summary, the analogue of cosmological particle creation should be 
realizable experimentally with present-day technology (in fact, the 
experimental efforts are already under way \cite{ion-trap}).
Unfortunately, simulating the other quantum effects with this set-up 
seems to be more complicated.
The main reason for these difficulties lies in the restricted 
parameter range (e.g., number and distance of ions) for which the 
coherent control can be maintained. 
As mentioned before, the advantages of this set-up lies in the 
ability to cool very close to the (motional plus electronic) 
ground state (better than 90\%) and to detect single phonons. 

\subsection{Bose-Einstein Condensates}

Another laboratory system capable of achieving very low temperatures 
experimentally is a Bose-Einstein condensate \cite{Dalfovo+Leggett}. 
In the dilute-gas limit, it can be described by the many-particle 
Hamiltonian density 
\bea
\label{heisen}
\hat{\cal H}
=
\hat\Psi^\dagger
\left(-\frac{\na^2}{2m}+V_{\rm ext}(t,\f{r})+
\frac{g}{2}\,\hat\Psi^\dagger\hat\Psi\right)\hat\Psi
\,,
\ea
where $m$ is the mass of the condensed particles (e.g., atoms) and 
$V_{\rm ext}(t,\f{r})$ the external trap potential.
The coupling $g>0$ describes the repulsion between the particles 
(in $s$-wave scattering approximation) and generates the internal 
pressure $p$. 
For dilute condensates consisting of many particles, the bosonic 
field operator can be approximately replaced by a c-number 
$\hat\Psi\approx\psi$ at low temperatures.
The order parameter $\psi$ corresponds to the macroscopically 
occupied wave function of the condensate and obeys the 
Gross-Pitaevski\v\i\/ equation 
\bea
\label{GP}
i\dot\psi=\left(-\frac{\na^2}{2m}+V_{\rm ext}(t,\f{r})+g|\psi|^2\right)\psi
\,.
\ea
Inserting the Madelung split 
$\psi(t,\f{r})=\sqrt{\varrho(t,\f{r})}\,e^{iS(t,\fk{r})}$
into density $\varrho$ and phase $S$, we obtain the equation of 
continuity $\dot\varrho+\na\cdot(\varrho\vau)=0$ for the 
condensate velocity $\vau=\na S/m$ and the Hamilton-Jacobi 
equation
\bea
\label{bernoulli}
\dot S+V_{\rm ext}+g\varrho+\frac{(\na S)^2}{2m}=
\frac{1}{2m}\,\frac{\na^2\sqrt{\varrho}}{\sqrt{\varrho}}
\,,
\ea
strongly resembles the Bernoulli equation apart from the term on the 
r.h.s. 
Indeed, on length scales much larger than the healing length 
$\xi=1/\sqrt{mg\varrho}$, this term can be neglected and we obtain 
a fluid dynamic description at distances $\gg\xi$. 

Therefore, the condensate can be described by an irrotational fluid. 
Vorticity may only occur via ``drilling a hole'' in the condensate,
i.e., a vortex at which center the condensate density vanishes.
Going around the vortex, the phase of the wave-function $\psi$ 
changes by an integer multiple of $2\pi$ due to the single-valued 
character of the wave-function $\psi$. 
Therefore, the circulation is quantized according to 
$\Gamma=\oint d\f{r}\cdot\vau\in2\pi{\mathbb N}/m$. 
On very large scales, however, a collection of many vortices seems to 
approach the classical Euler equation with rotation and even viscisity 
(shedding of phonons and vortex rings) including phenomena like 
turbulence (vortex tangle). 

The equation of state can be inferred from (\ref{bernoulli}) to be 
$p(\varrho)=g\varrho^2/(2m)$. 
Hence the sound velocity is given by $c^2_{\rm s}=g\varrho/m$ and 
the sound dispersion relation can be obtained by linearizing 
(\ref{bernoulli})
\bea
\label{disp-xi}
\left(\omega-\vau_0\cdot\f{k}\right)^2=
c^2_{\rm s}\f{k}^2+\left(\frac{\f{k}^2}{2m}\right)^2=
c^2_{\rm s}\f{k}^2\left(1+\frac{\xi^2}{4}\f{k}^2\right)
\,.
\ea
Thus, for dilute condensates, deviations from fluid dynamics 
at short wavelengths $\lambda\leq\ord(\xi)$ manifest themselves 
in a departure of the dispersion relation from the phonon branch 
$c_{\rm s}|\f{k}|$ towards the free-particle behavior 
$\f{k}^2/(2m)$. 
Apart from our good theoretical understanding of Bose-Einstein 
condensates (which is far less developed for most other fluids),
they offer a high degree of experimental controllability: 
The external potential $V_{\rm ext}$ can be generated by laser 
beams and thus there are no walls \cite{pessi}
which might induce friction etc. 
One may even realize effectively lower-dimensional condensates 
(if the transversal length scales are far below the healing 
length $\xi$). 
Furthermore, Bose-Einstein condensates are coupled only very 
weakly to the environment, which allow us to reach extremely 
low temperatures (of order nano-Kelvin).
Finally, for non-trivial topologies, one may generate persistent 
currents via trapping flux quanta imprinted on phase $S$.

Both, our good theoretical understanding and the high degree of 
experimental controllability, render Bose-Einstein condensates an 
ideal play-ground for theoretical and hopefully experimental 
investigations of the analogy to gravity.
In fact, most of the studies so far have been devoted to this system:  
In time-of-flight experiments, one uses an expansion of the condensate 
cloud in order to spatially amplify the desired signal.
On the classical level, the expansion can be described in terms of a
scaling parameter similar to Eq.~(\ref{scaling}).
The quantum fluctuations of the phonons, however, should imply effects 
analogous to cosmological particle creation resulting in frozen density 
(on the percent-level)
variations similar to cosmic inflation \cite{Expanding,BEC-cosmo}. 
Alternatively, expanding or contracting universes can be modelled 
\cite{qpt,two-component,BEC-cosmo} by temporal variations of the 
coupling $g$ (e.g., via a Feshbach resonance).
It is even possible to simulate a change of the signature of the 
metric by switching from repulsive to attractive interactions \cite{White} 
(which leads to a ``Bose-nova'' instability \cite{bose-nova}). 
For these effects, the healing length $\xi$ in Eq.~(\ref{disp-xi}) 
provides a natural UV cut-off and allows us to study its impact on 
the spectrum of produced particles \cite{Weinfurtner}. 

By inserting an atomic quantum dot as a detector model into the 
condensate, it is also possible to simulate the Gibbons-Hawking 
effect \cite{Fedichev} in a condensate in one spatial dimension
(remember that the scalar field is conformally invariant in 1+1 
dimensions). 
Going a step further and moving the atomic quantum dot through the 
(otherwise static) condensate with a varying velocity, one may pick 
up excitations due to non-adiabatic response of the quantum 
fluctuations \cite{Retzker}. 
This quantum effect is a bit similar to the Unruh effect, 
but there are important differences: 
The non-adiabatic response is caused by the Doppler shift 
(and hence works in the desired way in a certain direction only) 
and not by a real Lorentz time dilatation an in (\ref{Lorentz}).  

Classical effects such as superradiance, the white-hole instability, 
and the black-hole laser have been observed numerically by simulating 
the Gross-Pitaevski\v\i\/ equation (\ref{GP}) and should also be 
accessible experimentally \cite{Takeuchi}. 
For example, a multiply quantized vortex (which is known to be unstable)
is surrounded by the analogue of an ergo-region as in the Kerr metric 
and superradiance-like effects should play a role in its decay. 

In a two-component Bose-Einstein condensate, it is possible to mix the 
two species via optical transitions and thus to simulate a massive 
Klein-Fock-Gordon equation \cite{Massive}
via effectively breaking the $U(1)\times U(1)$ symmetry down to $U(1)$. 
This construction provides an example for the possible interplay 
between the cut-off scale and the mass scale \cite{Naturalness}. 

Finally, for Hawking radiation in Bose-Einstein condensates \cite{Garay}, 
the characteristic temperatures can be estimated from (\ref{gradient}).
Since the curved space-time analogy breaks down at length scales smaller 
than the healing length (typically below a micrometer) and the speed 
of sound is of order mm/s, we get a maximum Hawking temperature of order 
nano-Kelvin \cite{Savage}. 
The detection of the small number of emitted phonons could in principle 
\cite{Raizen} be done with similar methods as described in Sec.~\ref{Ions}, 
cf.~Fig.~\ref{lambda}, but for Bose-Einstein condensates, this technology 
is not as far advanced as for ion traps. 
Another interesting idea for detecting the Hawking process is to look 
at density-density correlations across the horizon \cite{Balbinot}, 
which show characteristic peaks due to the entanglement between the 
Hawking radiation and their partner particles described by a multi-mode 
squeezed state as in (\ref{multi-mode}). 
However, inserting typical values, these peaks are very weak and thus 
extremely hard to observe. 

Unfortunately, besides the aforementioned advantages, Bose-Einstein 
condensates go along with a major draw-back: 
All the gaseous condensates realized in the laboratory are only 
meta-stable states -- the true ground state is a solid. 
The main decay channels of the gaseous state are (inelastic)  
three-body collisions. 
Such an event transforms three indistinguishable bosons from the 
cloud into a molecule plus a remaining boson which carries away 
the large excess energy/momentum -- and thereby all three of them 
are effectively extracted from the condensate 
(formed of low-energy bosons).
Now, if all three of these bosons stem from the same macroscopically 
occupied single-particle wave-function $\psi$ of the condensate 
($\f{k}=0$), three-body collisions would just slowly diminish the 
number of condensed bosons.
However, in the presence of inter-particle interactions $g>0$ 
(which are necessary for the propagation of sound), the many-particle 
ground state does also contain a small population of the higher 
single-particle states $\f{k}>0$.
This small fraction is called quantum depletion since it is 
generated by the quantum fluctuations (plus the interaction $g$). 
Thus, if one of the three bosons involved in the inelastic collision 
stems from the quantum depletion and is removed, this event causes a 
deviation from the many-particle ground state -- i.e., an excitation 
\cite{Dziarmaga+Sacha}. 
Ergo, three-body collisions do also heat up the condensate, 
which might swamp the quantum signal to be detected \cite{Wuester}. 
For example, considering a Bose-Einstein condensate containing $10^7$
particles and $1\%$ quantum depletion with $1\%$ three-body losses 
(in total) during the experiment (which are already quite optimistic
values), there would be $\ord(10^3)$ noise phonons in addition to the
weak Hawking signal.
In order to avoid this problem (and other issues, such as the black hole 
laser instability), it is probably desirable to employ a  very fast 
detection method. 

\subsection{Surface Waves}\label{Surface}

As mentioned in Sec.~\ref{Further}, the analogy to gravity is not limited to
phonons but may apply to other quasi-particles as well.
As one example, the propagation of surface waves (ripplons) on top of 
locally irrotational flowing liquids corresponds to an effective metric 
very similar to Eq.~(\ref{PGL}) at long distances. 
This includes gravity waves (not to be confused with gravitational waves) 
in flowing water as well as third sound (i.e., surface waves) in liquid 
Helium (where gravity is replaced by van-der-Waals forces).  

As a result, the vortex flow which occurs when draining the bath tub,
for example, exhibits similarities to a rotating black hole and some 
of the associated instabilities might be related to superradiance.
As another scenario which can be realized in a sink, we may let the 
water jet from the tab impinge perpendicularly onto a flat surface 
(e.g., the sink bottom) and thereby generate a radial diverging flow. 
For suitable parameters, the radial flow is faster than the surface 
wave speed (i.e., ``superluminal'') in the middle and becomes slower 
(``subluminal'') at some given radius.
This point corresponds to a white-hole horizon \cite{hydraulic-jump} 
and is known as hydraulic jump in view of the instabilities occurring 
there: the quasi-regular radial flow in the middle becomes irregular 
beyond that radius and the water height changes abruptly (jump). 
The instabilities generating this phenomenon are probably related to 
the Eardley effect. 

A similar experiment has been carried out in a water tank 
\cite{Rousseaux} which was quite large (in order to minimize viscosity
effects etc.) and allowed the quantitative determination of the mixing 
of positive $\phi_{\rm out}^+(\omega)$ and negative 
$\phi_{\rm out}^-(\omega)$ frequency modes 
$\alpha_\omega\phi_{\rm out}^+(\omega)+\beta_\omega\phi_{\rm out}^-(\omega)
\leftrightarrow\phi_{\rm in}(\omega)$
roughly analogous to the time-reversed Hawking process,  
cf.~Sec.~\ref{Hawking}. 
However, the measured results $\alpha_\omega$ and $\beta_\omega$ 
did not match the expected values and the curved space-time analogy
probably does not apply quantitatively to this experiment. 

In summary, the simulation of classical effects with surface waves is 
potentially difficult but in principle possible -- while it is hard to 
see how analogues of quantum phenomena could be detected. 

\subsection{Optical Fibers and Wave-Guides}

In order to model a black hole horizon for electromagnetic waves in the 
same way as in the previous sections, the velocity of the medium should 
exceed the speed of light in that medium \cite{Soff} at some point, 
cf.~Eq.~(\ref{gordon-o}). 
Unless one uses meta-materials or other slow-light media 
(which induce further problems \cite{Slow}), this is very difficult 
to realize experimentally.
Therefore, one may chose an alternative route:
Instead of actually moving the medium, one could send a pulse 
(representing a phase boundary) through the medium such that the speed 
of light in front of the pulse is larger than its velocity of propagation 
-- whereas the speed of light behind the pulse is smaller. 
Transforming to the inertial reference fame co-moving with the pulse,
we see that it also corresponds to a black hole horizon.

This scenario facilitates much larger velocities and hence higher 
Hawking temperatures, cf.~Eq.~(\ref{gradient}).
For electromagnetic wave-guides \cite{wave-guide} supporting radio 
or micro waves, the Hawking temperature could range up to fractions 
of a Kelvin.
For optical fibers \cite{Philbin}, the value could be even larger.
However, for normal non-linear optical media
(where the refractive index increases with intensity), 
the Kerr effects tends to make the front end of a light pulse
(i.e., the black hole horizon) flat and its rear end 
(i.e., the white hole horizon) steep.
As a result, the effective surface gravity for the black hole horizon
(and hence the Hawking temperature) becomes rather small while the 
effective surface gravity for the white hole horizon is increased. 
For this reason, the recent experiment in optical fibers \cite{Philbin}
considered the white hole horizon and measured the classical mixing 
of positive $\phi_{\rm out}^+(\omega)$ and negative 
$\phi_{\rm out}^-(\omega)$ frequency modes 
$\alpha_\omega\phi_{\rm out}^+(\omega)+\beta_\omega\phi_{\rm out}^-(\omega)
\leftrightarrow\phi_{\rm in}(\omega)$
roughly analogous to the time-reversed Hawking process,  
cf.~Sec.~\ref{Hawking}. 

One of the main advantages of these electromagnetic scenarios 
lies in the fact that the amplification and detection of photons 
(in the relevant energy range) is much easier than in the case of phonons.  
Nevertheless, observing quantum effects is still quite hard because the 
small signal to be detected arrives just before the huge pulse 
(in the black hole case) or, even worse, just after the pulse 
(for white hole analogues).  

\subsection{Ultra-intense Lasers and the Unruh Effect}

In all of the previous example, the speed of light in vacuum has been 
effectively replaced by the reduced propagation velocity of 
quasi-particles (such as phonons, ripplons, or photons) in a medium. 
In the following, we shall discuss a scenario where this is not the case 
and consider a real relativistic quantum effect.
With modern lasers, it is possible to reach  extremely high intensities 
via suitable focussing techniques \cite{Mourou}. 
Electrons under the influence of these strong electric fields would 
undergo an immense acceleration and thus experience the Minkowski vacuum 
as a thermal state with a relatively large temperature \cite{Chen+Tajima}. 

However, the electron is not a good photon detector since it cannot 
absorb a photon while passing to an excited internal state -- and thus 
directly observing the Unruh effect is rather difficult. 
On the other hand, the electron {\em can} scatter photons from one into 
another mode via Thomson (or Compton) scattering.
Consequently, switching to the accelerated frame co-moving with the 
electron, there is a finite probability that the electron scatters a 
(``virtual'') photon out of the thermal bath into another mode. 
In terms of the squeezed state in Eq.~(\ref{multi-mode}), this corresponds 
to $\hat a^\dagger_{\rm R,\,right}(\f{k})\hat a_{\rm R,\,right}(\f{k'})
\ket{0_{\rm Min}}$ with $\omega(\f{k})=\omega(\f{k'})$ due to energy 
conservation (in the accelerated frame).
Translation back into the inertial frame yields the creation of an 
entangled {\em pair} of photons out of the quantum vacuum fluctuations 
due to the non-inertial scatterer.
This result can be understood in the following way:
Since the Minkowski vacuum is annihilated by a linear combination 
of Rindler operators in the left and right wedges \cite{Unruh-prd}
\bea
\left(
e^{-\pi\omega/a}\,
\hat a^\dagger_{\rm R,\,left}(\omega)+
\hat a_{\rm R,\,right}(\omega)
\right)\ket{0_{\rm Min}}=0
\,,
\ea
removing one Rindler particle 
$\hat a_{\rm R,\,right}(\f{k'})\ket{0_{\rm Min}}$
from the right wedge (e.g., via absorption by the accelerated detector) 
leaves its partner particle in the left wedge behind 
$\hat a^\dagger_{\rm R,\,left}(\omega)\ket{0_{\rm Min}}$
and thereby liberates it to become a real excitation \cite{Happens}. 
Furthermore, inserting an additional particle into the right wedge 
$\hat a^\dagger_{\rm R,\,right}(\f{k})\ket{0_{\rm Min}}$ does also 
cause a deviation from the vacuum state $\ket{0_{\rm Min}}$ and thus
corresponds to a real excitation, i.e., photon.

The created photon pairs may then serve as a signature of the Unruh 
effect and their entanglement (e.g., in polarization \cite{Habs}) 
reflects the correlations between the left and right Rindler wedge,
cf.~Eq.~(\ref{multi-mode}).
As stated before, these correlations across the horizon are genuine 
features of the considered quantum effects such as Hawking radiation. 
The two-photon nature of these signatures of the Unruh effect should
(at least in principle) allows us to distinguish it from other effects 
such as classical Larmor radiation (given off by all accelerated charges). 

For a close analogy to the Unruh effect (uniform acceleration),
one would like to have a very strong and approximately constant electric 
field and a negligible magnetic field.
However, it is rather hard to obtain a strong enough quantum signal in 
this set-up and to distinguish it from the classical background. 
A stronger signal and better discrimination from Larmor radiation could 
be achieved via an alternative scenario, which does, however, no longer
correspond to uniform acceleration:
Colliding an ultra-relativistic electron beam with a counter-propagating 
laser pulse, the electric field felt by the electrons and hence their 
acceleration is strongly amplified by the Lorentz boost factor. 
Furthermore, the quantum (Unruh) signal and the classical (Larmor) 
background are better separated in phase space (energy and momentum) 
in this situation \cite{Habs}.

Finally, it should be mentioned that there are also further indirect 
signatures of the Unruh effect \cite{Rosu}, some of them have already 
been observed.
For example, the residual polarization variance of electrons in accelerator 
rings can partly be attributed to the Unruh effect \cite{orbit}: 
The orbiting electron is constantly (though not uniformly) accelerated 
and thus experiences the Minkowski vacuum as a nearly thermal bath, 
which inhibits a perfect polarization \cite{orbit}. 
Another example are accelerated atoms in a cavity \cite{Scully}.
However, in this scenario, the main signal is not generated by the 
actual acceleration of the atom, but by non-adiabatic switching 
effects. 
Therefore, this set-up displays more similarities to the ion-trap 
configuration in Sec.~\ref{Ions}. 

\subsection{Schwinger Mechanism in the Laboratory?}

With the laser systems currently under preparation \cite{ELI}, 
it should become possible to approach the Schwinger limit $E_S$ 
within two orders of magnitude in the focus of the high-intensity laser.  
Via high-harmonic focusing \cite{Pukhov}, it may even be possible to 
reach much higher intensities. 
Therefore, the experimental observation of the Schwinger mechanism, 
which is a non-perturbative QED vacuum effect, may become within reach 
in the near future. 
Note that its non-perturbative character clearly distinguishes the 
Schwinger mechanism from perturbative (multi-photon) pair-creation 
effects, which have already been observed in accelerators.  

However, since it is probably rather hard to actually reach the 
Schwinger limit $E_S$ and the pair-creation rate is exponentially 
(non-perturbatively) suppressed $\propto\exp\{-\pi E_S/E\}$, 
it is desirable to increase the signal.
As one possibility, one could superimpose the strong and slow electric 
field with weak and fast electromagnetic wiggles of frequency 
$\Omega<m$.
Even though these wiggles cannot produce electron-positron pairs 
directly, they help the ``virtual'' electrons to penetrate into the 
gap and thus decrease the distance between the classical turning
points $x_\pm=\pm(m-\Omega)/(qE)$.
As a result, the tunnelling exponent in Eq.~(\ref{exponent}) 
is altered to 
\bea
\int\limits_{x_-}^{x_+} dx\,\sqrt{m^2-(qEx-\Omega)^2}
=
\frac{m^2}{4qE}\times 
f\left(\frac{\Omega}{m}\right)\,,
\ea
with $f(\chi)=\pi+2\arcsin(1-2\chi)+4\sqrt{\chi}(1-2\chi)\sqrt{1-\chi}$,
which can be approximated by $f(\chi)\approx2\pi(1-\chi)$ in the relevant 
interval $\chi\in[0,1]$.
Consequently, the tunnelling exponent can be decreased by the 
weak and fast electromagnetic wiggles \cite{Dunne},
e.g., for $\chi=1/2$, 
the pair-creation rate scales with $\propto\exp\{-\pi E_S/(2E)\}$, 
which is a huge enhancement for $E\ll E_S$. 

\section{The Big Picture}\label{big}

Now, after having discussed various scenarios for recreating fundamental
effect is the laboratory, let us turn to the question of what we can learn 
form these considerations. 
First and foremost, these scenarios facilitate (at least in principle) 
an experimental test of the striking predictions of quantum field theory 
in non-trivial backgrounds. 
This allows us to test the methods, assumptions, and approximations 
underlying these predictions (cf.~the trans-Planckian problem discussed 
in Sec.~\ref{Hawking}, for example). 
Furthermore, after detecting the desired effect itself, we might 
experimentally study corrections induced by interactions between 
the quasi-particles (which are theoretically not fully understood)
as well as non-perturbative aspects (e.g., the Schwinger mechanism). 

Apart from these experimental issues, the study of the laboratory 
analogues does also provide new ideas for investigating the possible 
microscopic structures of quantum gravity inspired by condensed matter.
For example, the trans-Planckian issue can be approached via considering 
a change in the dispersion relation, cf.~Sec.~\ref{Hawking}. 
These calculations suggest that (in the absence of a ``UV-catastrophe'') 
Hawking radiation is basically a low-energy process.  
This observation casts some doubt on the frequently suggested resolution 
of the black hole information ``paradox'' postulating subtle correlations 
in the outgoing Hawking radiation which carry away the information.  
If the Hawking particles are created at low energies near the horizon, 
it is hard to see how they can be influenced by the information ``stored''  
in the black hole, which is presumably located near the singularity. 

As another important point, the laboratory analogues constitute nice
examples for the distinction between universal (emergent) phenomena 
and system-specific (microscopic) features, which is also crucial for 
our understanding of quantum gravity.
While an effective metric, i.e., an effective Lorentz invariance, 
seems to emerge naturally in many condensed matter systems, the 
principle of equivalence and the Einstein equations require some 
additional input (kinematic versus dynamics).
The combination of kinematic and dynamics becomes relevant for the 
back-reaction of quantum fluctuations onto the approximately classical 
background. 
Studying Bose-Einstein condensates as an example, one finds that the 
naive sum of the zero-point energy/pressure of the phonon modes up to 
some cut-off yields a completely wrong result \cite{back-reaction}. 
This lesson should be relevant for the cosmological constant problem,
for example.  

The analogy to condensed matter does also teach us that quantizing a 
classical macroscopic theory is not always working (even though the 
quantization of the linearized quasi-particles might work). 
In the case of electrodynamics, the direct transition from classical 
to quantum description was very successful.
For fluid dynamics, however, this procedure is very problematic.  
For example, starting from the classical Euler equation, one does not 
get the quantization of vorticity right. 
In the case of superfluids, for instance, the circulation quantum depends 
on the mass of the condensed particles, which does not appear at all in 
the classical Euler equation.
The main question, then, is whether gravity is more similar to 
electrodynamics (where a direct quantization works very well) 
or to fluid dynamics (where the correct quantization requires some 
knowledge about the microscopic structure).
The underlying symmetries, the complex constraint structure, and the 
UV-divergences resulting from a naive quantization seem to point towards 
the latter -- but this question must be answered in the future. 

Finally, the analogy between condensed matter and gravity does also
allow us to understand laboratory physics from a different point of view,
i.e., in terms of the universal geometrical concepts (e.g., horizons)
known from gravity.  
For low energies, we may forget the underlying microscopic structure and 
consider the effective metric only. 
As a result, very different condensed matter systems -- such as an 
expanding fluid on the one hand and a medium at rest with a time-dependent 
speed of sound on the other hand -- may exhibit basically the same physical
effects (e.g., cosmological particle creation) due to the coordinate 
invariance of the effective metric. 

\section*{Acknowledgements}

This work was supported by the Emmy-Noether Programme of the German
Research Foundation (DFG) under grant \# SCHU~1557/1. 
The author acknowledges fruitful discussions with Bill Unruh and 
many others (who cannot be listed here due to the irreconcilability 
of doing justice to everyone of the one hand and space restrictions 
on the other hand.) 


\end{document}